# A relation between conditional entropy and conditional expectation, to evaluate secrecy systems

Thibault de Valroger [(*)]

Abstract

We demonstrate an intuitive relation between conditional entropy and conditional expectation that is useful when one want to compare them as measurement tools to evaluate secrecy systems. In particular, we give a Security Property applicable to general vector variables in $\mathbb{R}^n$, using measurement based on vector quadratic distance, and we show that one can derive variables that can be measured with Csiszàr and Körner secrecy capacity measurement [2], based on conditional entropy, with conserving the same order relation.

## I. Introduction

When considering communication systems in which three entities $A$, $B$ and $E$ receive respectively a random variable $X$, $Y$, and $Z$, it has been an important domain of research to determine under what conditions on the variables $X$, $Y$, and $Z$ the communication system is Information Theoretically secure, meaning that it enables the legitimate partners $A$ and $B$ to exchange data that remain hidden to the eavesdropping opponent $E$, even if $E$ has unlimited computation and storage power.

The first necessary achievement in order to obtain Information Theoretical Security is called Advantage Distillation. It can be intuitively understood as the fact that $A$ and $B$ obtain variables $X$ and $Y$ that are 'closer' than say $X$ and $Z$, according to some distance over the sample space.

Csiszàr and Körner have introduced in [2] a measurement, based on conditional entropy, that lower-bounds secrecy capacity $C_s$, meaning the maximum rate at which $A$ and $B$ can exchange Information Theoretically secure bits. That lower-bound is:

$$C_s \geq \max_{P_X}\bigl(H(X|Z) - H(X|Y)\bigr)$$

On the other hand, entropy is not an easy function to manipulate in many contexts, and it appears more natural, typically in the Euclidean space $\mathbb{R}^n$, to measure Advantage between the partners $A$ and $B$ over the opponent $E$ by using the quadratic distance in $\mathbb{R}^n$.

This article gives an equivalence between a measurement of Advantage based on quadratic distance and the Csiszàr and Körner's measurement of Advantage applied to a transformed version of the variables $X$, $Y$, and $Z$.

*(\*) See contact and information about the author at last page*

## II. Presentation and proof of the general result

We consider $\|\cdot\|$ the quadratic distance over the Euclidean space $\mathbb{R}^n$. For $X \in \mathbb{R}^n$:

$$\|X\|^2 \triangleq \sum_{s=1}^{n} x_s^2$$

The technical result that we will demonstrate is the following:

**Lemma 1.**

> For $X, Y, Z \in \mathbb{R}^n$, there exist (i) a set $E$, (ii) an independent random variable $V$ taking values in $E$, (iii) a non trivial pre-condition $\mathcal{C}_X$ and (iv) a function $f: \mathbb{R}^n \times E \to \{0,1\}$, such that:

$$\|X - Z\| \geq \|X - Y\| \Leftrightarrow E[f(X,V) \oplus f(Z,V)|X,Z,\mathcal{C}_X,\mathcal{C}_Z] \geq E[f(X,V) \oplus f(Y,V)|X,Y,\mathcal{C}_X,\mathcal{C}_Y]$$

Proof of Lemma 1 can be found in Annex 1.

We thus apply the randomized transformation:

$$X^* = f(X, \Theta)$$

$$Y^* = f(Y, \Theta)$$

$$Z^* = f(Z, \Theta)$$

It has been shown by Wyner [3] that, for binary random variable $X^*, Y^*, Z^*$ (typically being the received bit flows of $A, B, E$ in a secrecy system), provided that $P(X^* \neq Y^*) < 1/2$:

$$P(X^* \neq Y^*) < P(X^* \neq Z^*) \Longrightarrow H(X^*|Z^*) > H(X^*|Y^*)$$

Therefore, by noticing that $P(X^* \neq Y^*) = E[X^* \oplus Y^*]$, we can explicit that in the conditions above, we have the relation:

$$E[\|X - Z\|^2] > E[\|X - Y\|^2] \Longrightarrow H(X^*|Z^*) > H(X^*|Y^*)$$

## III. Interest of the result

In a secrecy system, the passive eavesdropping opponent has access only to a set of public information $J$, and then the optimal estimation of $X$ by an observer who only knows $J$ is of course given by

$$E[X|J]$$

This variable minimizes

$$\inf_{Z(J)} E[\|X - Z(J)\|^2]$$

It is therefore a consequence of the above relation that, if the observer $E$ does not have access to the optimal estimation variable $E[X|J]$ for any reason governed by the Security Model, then we may have:

$$E[\|X - Z(J)\|^2] > E[\|X - E[X|J]\|^2] \tag{I}$$

for any strategy $Z(J)$ chosen by the opponent, and thus:

$$H(X^*|Z^*) > H(X^*|J)$$

An example of such Security Model where the observer $E$ does not have access to the optimal estimation variable $E[X|J]$ is Deep Random secrecy [5], in which it is shown that the absence of full knowledge of probability distribution of $X$ by the opponent $E$ may be exploited to ensure that the inequality $(I)$ is strict, thus enabling to obtain strictly positive Secrecy Capacity $C_S$. Note however that in a secrecy system, the legitimate partner $B$ may neither have access to an estimation as accurate as the optimal estimation $E[X|J]$, and therefore, when analyzing a secrecy system, the inequality $(I)$ is not alone a guarantee of security.

In general, it is useful to be able to study Advantage Distillation with to a tool like quadratic distance, that benefits from solid mathematical framework, rather than directly conditional entropy.

**Who is the author ?**
I have been an engineer in computer science for 20 years. My professional activities in private sector are related to IT Security and digital trust, but have no relation with my personal research activity in the domain of cryptology. If you are interested in the topics introduced in this article, please feel free to establish first contact at tdevalroger@gmail.com

## Annex – Proof of Lemma 1

■ Let's first establish the result for $n = 1$:

We denote $\Gamma(x) = [x] \mod 2$, in which $[x] \triangleq \max_{u \in \mathbb{Z}}\{u \leq x\}$

For $X, Y \in [-1/2, 1/2[$ (which represents $\mathcal{C}_X$ and $\mathcal{C}_Y$), and $V \in [0,1[$ we have:

$$E[\Gamma(X - V) \oplus \Gamma(Y - V) | X, Y] = |X - Y|$$

and thus the equivalence is immediate.

The proof of the result for $n \geq 2$ is the following:

Let's assume that $\|X\| = \varepsilon$ is a public condition $\mathcal{C}_X$, then it is natural to impose as well $\mathcal{C}_Y$ and $\mathcal{C}_Z$. We will denote by $I_p$ the identity matrix of size $p$. For $p \leq n - 2$, we consider the isometries:

$$U_p(\theta) \triangleq \begin{pmatrix} I_p & \cdots & & & 0 \\ \vdots & \cos\theta & \sin\theta & & \vdots \\ & -\sin\theta & \cos\theta & & \\ 0 & & \cdots & & I_{n-p-2} \end{pmatrix}$$

For $\Theta = (\theta_1, \ldots, \theta_{n-1}) \in [0, 2\pi[^{n-1}$, we consider the functions:

$$U(\Theta) \triangleq \prod_{p=1}^{n-1} U_p(\theta_p)$$

$$F(X, \Theta) \triangleq U(\Theta) X$$

$$f(X, \Theta) \triangleq \bigoplus_{m=1}^{n} ([F(X, \Theta) \cdot e_m] \mod 2)$$

By recurrence, we get:

$$U(\Theta) e_n = \begin{pmatrix} \prod_{p=1}^{n-1} \sin\theta_p \\ \vdots \\ \cos\theta_{j-1} \prod_{p=j}^{n-1} \sin\theta_p, \quad 2 \leq j \leq n-1 \\ \vdots \\ \cos\theta_{n-1} \end{pmatrix}$$

which implies that $F(\varepsilon e_n, \Theta)$ is a surjection from $[0, 2\pi[^{n-1}$ on the n-sphere $\varepsilon S_{n-1}$.

Let's define the equivalence relation $\mathcal{R}_X$ in $[0, 2\pi[^{n-1}$:

$$\Theta \mathcal{R}_X \Theta' \iff F(X, \Theta) = F(X, \Theta')$$

The equivalence classes are denoted $\mathcal{C}_X(U(\Theta) X)$. Each equivalence class $\mathcal{C}_{e_n}(U(\Theta) e_n)$ has exactly $2^{n-2}$ elements. If $\Theta_X$ is an element of $\mathcal{C}_{e_n}(X)$, then:

$$X = U(\Theta_X)e_n$$

$$X' = U(\Theta_{X'})e_n = U(\Theta_{X'})^t U(\Theta_X)X = U\big((\Theta_{X'} - \Theta_X) \bmod 2\pi\big)X$$

which implies that, not only for $\varepsilon e_n$ but for any $X \in \varepsilon S_{n-1}$ we have:

(i) $F(X, \Theta)$ is a surjection from $[0, 2\pi[^{n-1}$ on the n-sphere $\varepsilon S_{n-1}$
(ii) $\forall X, X' \in \varepsilon S_{n-1}$ all classes $\mathcal{C}_X(X')$ contains also exactly $2^{n-2}$

$$\begin{array}{ccc}
[0, 2\pi[^{n-1} & & \\
\downarrow s_{\mathcal{R}_X} & \searrow F(X, \Theta) & \\
[0, 2\pi[^{n-1}/\mathcal{R}_X & \xrightarrow{\tilde{F}} & \varepsilon S_{n-1}
\end{array}$$

By decomposing $F$ canonically as shown above, we obtain that $\tilde{F}$ is a bijection and courses uniformly the n-sphere $\varepsilon S_{n-1}$ when $(\theta_0, \ldots, \theta_{n-2})$ courses uniformly $[0, 2\pi[^{n-1}/\mathcal{R}_X$. It implies that:

$$E[f(X, \Theta) \oplus f(Y, \Theta) | X, Y, \mathcal{C}_X, \mathcal{C}_Y] = \varphi(\|X - Y\|)$$

only depends on $\|X - Y\|$. $\bigoplus_{m=1}^{n} ([(\cdot) \cdot e_m] \bmod 2)$ corresponds to the mapping of a vector in the n-dimensional checkboard where each unit hypercube 'colored in white' has immediate adjacent unit hypercube only 'colored in black'.

Then, it is clear that $\varphi(0) = 0$, and $\varphi(x) > 0$ for $x > 0$. It implies that $\varphi$ is locally increasing for $x \le \varepsilon$, if $\varepsilon$ is chosen sufficiently small. ∎